\def\etal{{\frenchspacing\it et al.}}

\def\beq#1{\begin{equation}\label{#1}}
\def\eeq{\end{equation}}
\def\beqa#1{\begin{eqnarray}\label{#1}}
\def\eeqa{\end{eqnarray}}

\def\la{\mathrel{\mathpalette\fun <}}

\def\fun#1#2{\lower3.6pt\vbox{\baselineskip0pt\lineskip.9pt
        \ialign{$\mathsurround=0pt#1\hfill##\hfil$\crcr#2\crcr\sim\crcr}}}
        
\def\xi{{{\bf x}^b}}

\documentclass[twocolumn,aps,showpacs,showkeys,nofootinbib]{revtex4}
\usepackage{epsfig}

\newcommand{\be}{\begin{equation}}
\newcommand{\ee}{\end{equation}}
\newcommand{\ba}{\begin{eqnarray}}
\newcommand{\ea}{\end{eqnarray}}

\begin{document}
\input{epsf.sty}

\title{Distance Measurements from Supernovae and Dark Energy Constraints}
\author{Yun~Wang}
\address{Homer L. Dodge Department of Physics \& Astronomy, Univ. of Oklahoma,
                 440 W Brooks St., Norman, OK 73019;
                 email: wang@nhn.ou.edu}

                 \today

\begin{abstract}

Constraints on dark energy from current observational data are
sensitive to how distances are measured from Type Ia supernova 
(SN Ia) data. We find that flux-averaging of SNe Ia can be used
to test the presence of unknown systematic uncertainties, and
yield more robust distance measurements from SNe Ia.
We have applied this approach to the nearby+SDSS+ESSENCE+SNLS+HST 
set of 288 SNe Ia, and the ``Constitution''of set 397 SNe Ia. 
Combining the SN Ia data with cosmic microwave background anisotropy 
data from Wilkinson Microwave Anisotropy Probe five year observations, 
the Sloan Digital Sky Survey baryon acoustic oscillation measurements, 
the data of 69 gammay-ray bursts, and the Hubble constant measurement
from the Hubble Space Telescope project SHOES, we measure
the dark energy density function $X(z)\equiv \rho_X(z)/\rho_X(0)$ 
as a free function of redshift (assumed to be a constant at $z>1$
or $z>1.5$). Without flux-averaging of SNe Ia, 
the combined data using the ``Constitution'' set of SNe Ia seem to 
indicate a deviation from a cosmological constant at $\sim$95\% confidence 
level at $0\la z \la 0.8$; they are consistent with a cosmological constant 
at $\sim$68\% confidence level when SNe Ia are flux-averaged. The combined 
data using the nearby+SDSS+ESSENCE+SNLS+HST data set of SNe Ia are 
consistent with a cosmological constant at 68\% confidence level 
with or without flux-averaging of SNe Ia, and give dark energy constraints 
that are significantly more stringent than that using the ``Constitution'' 
set of SNe Ia. Assuming a flat universe, dark energy is detected at
$> 98$\% confidence level for $z\leq 0.75$ using the combined data
with 288 SNe Ia from nearby+SDSS+ESSENCE+SNLS+HST,
independent of the assumptions about $X(z\geq 1)$.
We quantify dark energy constraints without assuming a flat universe
using the dark energy Figure-of-Merit (FoM) for both $X(z)$ and a dark 
energy equation-of-state linear in the cosmic scale factor.

\end{abstract}



\maketitle


\section{Introduction}

The observational evidence of cosmic acceleration \cite{Riess98,Perl99}
continues to strengthen with time 
(e.g., \cite{Komatsu09,Sollerman09,Serra09,Reid09,Biswas09,Daly09,Samushia09,Shafieloo09}). 
However, we are still in the dark about the nature of the observed cosmic 
acceleration; whether it is due to an unknown energy component
\cite{Freese87,Linde87,Peebles88,Wett88,Frieman95,Caldwell98,Kaloper06,Chiba09},
i.e., dark energy, or a modification of general relativity 
\cite{SH98,Parker99,Boisseau00,DGP00,Freese02,Pad08,Kahya09,OCallaghan09}. For
recent reviews, see \cite{Copeland06,Ruiz07,Ratra07,Frieman08,Caldwell09,Uzan09}.

Differentiating dark energy from modified gravity as the
cause for the observed cosmic acceleration requires ambitious
observational projects of galaxy redshift surveys \cite{Cimatti09}
and weak lensing surveys \cite{Refregier09} that are still in the planning stages.
Currently available data allow us to begin to test whether
the observed cosmic acceleration could be due to a cosmological constant,
the simplest possibility for dark energy.

A critical challenge in our understanding of dark energy is
the control of known and unknown systematic uncertainties.
Here we investigate methods for measuring distances from Type Ia 
supernova (SN Ia) data, and their impact on dark energy constraints
from current observational data. 

In the usual likelihood analysis of distances to SNe Ia that
directly utilizes the measured brightness of each SN Ia,
the statistics is dominated by the SNe Ia with
the smallest measurement errors. This could lead to biased distance
measurements in the presence of unknown systematic errors.
We will show that this pitfall can be removed by flux-averaging
SN Ia data in redshift bins (which also reduces the bias in estimated 
parameters due to weak gravitational lensing of the highest redshift 
SNe Ia). 

We describe our method in Sec.II, present our results in Sec.III,
and conclude in Sec.IV.

\section{Method}
\label{sec:method}

The minimal way to include dark energy in the standard cosmological 
framework is to add a new energy component with density $\rho_X(z)$.
The Friedman equation becomes
\ba
\label{eq:H(z)}
&&H^2(z)  \equiv  \left(\frac{\dot{a}}{a}\right)^2 \\
 &= &H_0^2 \left[ \Omega_m (1+z)^3 +\Omega_r (1+z)^4 +\Omega_k (1+z)^2 
+ \Omega_X X(z) \right],\nonumber
\ea
where $\Omega_m+\Omega_r+\Omega_k+\Omega_X=1$, and
the dark energy density function $X(z)$ is defined as
\be
X(z) \equiv \frac{\rho_X(z)}{\rho_X(0)}.
\ee
Note that $\Omega_r \ll \Omega_m$, thus the $\Omega_r$ term
is usually omitted in dark energy studies, since dark energy
should only be important at late times.

The comoving distance to an object at redshift $z$ is given by:
\ba
\label{eq:r(z)}
 & &r(z)=cH_0^{-1}\, |\Omega_k|^{-1/2} {\rm sinn}[|\Omega_k|^{1/2}\, \Gamma(z)],\\
 & &\Gamma(z)=\int_0^z\frac{dz'}{E(z')}, \hskip 1cm E(z)=H(z)/H_0 \nonumber
\ea
where ${\rm sinn}(x)=\sin(x)$, $x$, $\sinh(x)$ for 
$\Omega_k<0$, $\Omega_k=0$, and $\Omega_k>0$ respectively.

\subsection{Analysis of SN Ia Data}

The published SN Ia data sets usually give the distance modulus
to each SN Ia (marginalized over calibration parameters):
\be
\label{eq:m-M}
\mu_0 \equiv m-M= 5 \log\left[\frac{d_L(z)}{\mathrm{Mpc}}\right]+25,
\ee
where the luminosity distance $d_L(z)=(1+z)\, r(z)$, with the comoving
distance $r(z)$ given by Eq.(\ref{eq:r(z)}).

We consider the two latest compilations of SN Ia data, the ``Constitution''
set of 397 SNe Ia by Hicken et al. (2009) \cite{Hicken09}, and the
nearby+SDSS+ESSENCE+SNLS+HST set of 288 SNe Ia by Kessler et al. (2009)
\cite{Kessler09}. The ``Constitution'' set used SALT \cite{Guy05} to
fit the SN Ia light curves. We use the nearby+SDSS+ESSENCE+SNLS+HST set
that used SALT2 \cite{Guy07} for SN Ia lightcurve-fitting. We do not
use the nearby+SDSS+ESSENCE+SNLS+HST set that used MLCS 
\cite{Phillips93,Riess95,Jha07} for SN Ia lightcurve-fitting, as the
MLCS method appears to introduce some systematic biases \cite{Kessler09}.

It has been noted that the ``Constitution'' set of SNe Ia 
(and other prior compilations of SNe Ia) appear to be inhomogeneous,
possibly due to unknown systematic effects \cite{Qi09,Sanchez09}.
Here we use the flux-averaging of SNe Ia to help reduce the impact of
unknown systematic effects.

Flux-averaging of SNe Ia was proposed to reduce the effect of the weak 
lensing of SNe Ia on cosmological parameter estimation \cite{Wang00b}. 
The basic idea is that because of flux conservation in gravitational
lensing, the average magnification of a large number of SNe Ia at
the same redshift should be unity. Thus averaging the observed flux from
a large number of SNe Ia at the same redshift can recover the unlensed
brightness of the SNe Ia at that redshift.

Wang \& Mukherjee (2004) \cite{WangPia04} and Wang (2005) 
\cite{Wang05} developed a consistent framework for flux-averaging SNe Ia. 
Appendix A of Wang \& Mukherjee (2007) \cite{WangPia07} describes 
in detail the recipe for flux-averaging SNe Ia used here; a public code
is available at http://www.nhn.ou.edu/$\sim$wang/SNcode/.
Here we do {\it not} marginalized over $H_0$, since it is absorbed in
the ``nuisance parameter'' $M_{SN}$ that is marginalized over
in our likelihood analysis.

Since the SNe Ia in each redshift bin are flux-averaged and 
{\it not} used directly in the likelihood analysis, the systematic
bias that results from unknown systematic errors from individual
SNe Ia can be minimized. For homogeneous data without unknown
systematic errors, flux-averaging should not change the results
qualitatively at intermediate redshifts (where the lensing
effect is negligible). Thus in addition to reducing lensing and 
lensing-like systematic effects, flux-averaging of SN Ia data provide 
a useful test for the presence of unknown systematic effects
at intermediate redshifts. We limit the flux-averaging to
SNe Ia at $z\geq 0.2$, since the SNe Ia at lower redshifts 
are better understood.

\subsection{CMB data}
\label{sec:CMB}

CMB data give us the comoving distance to the photon-decoupling surface 
$r(z_*)$, and the comoving sound horizon 
at photo-decoupling epoch $r_s(z_*)$ \cite{Page03}.
Wang \& Mukherjee 2007 \cite{WangPia07} showed that
the CMB shift parameters
\be
R \equiv \sqrt{\Omega_m H_0^2} \,r(z_*), \hskip 0.1in
l_a \equiv \pi r(z_*)/r_s(z_*),
\ee
together with $\omega_b\equiv \Omega_b h^2$, provide an efficient summary
of CMB data as far as dark energy constraints go.
Using $\{R,l_a,\omega_b\}$ is equivalent to using
$\{R,l_a,z_*\}$ as CMB distance priors \cite{Wang08a}. 

The comoving sound horizon at redshift $z$ is given by
\ba
\label{eq:rs}
r_s(z)  &= & \int_0^{t} \frac{c_s\, dt'}{a}
=cH_0^{-1}\int_{z}^{\infty} dz'\,
\frac{c_s}{E(z')}, \nonumber\\
 &= & cH_0^{-1} \int_0^{a} 
\frac{da'}{\sqrt{ 3(1+ \overline{R_b}\,a')\, {a'}^4 E^2(z')}},
\ea
where $a$ is the cosmic scale factor, $a =1/(1+z)$, and
$a^4 E^2(z)=\Omega_m (a+a_{\rm eq})+\Omega_k a^2 +\Omega_X X(z) a^4$,
with $a_{\rm eq}=\Omega_{\rm rad}/\Omega_m=1/(1+z_{\rm eq})$, and
$z_{\rm eq}=2.5\times 10^4 \Omega_m h^2 (T_{CMB}/2.7\,{\rm K})^{-4}$.
The sound speed is $c_s=1/\sqrt{3(1+\overline{R_b}\,a)}$,
with $\overline{R_b}\,a=3\rho_b/(4\rho_\gamma)$,
$\overline{R_b}=31500\Omega_bh^2(T_{CMB}/2.7\,{\rm K})^{-4}$.
We take $T_{CMB}=2.725$ following Komatsu et al. (2009) \cite{Komatsu09}, 
since we will use the CMB bounds derived by them.

Here we use the covariance matrix of [$R(z_*), l_a(z_*), z_*, r_s(z_d)]$ from
the five year WMAP data \cite{Komatsu09,Komatsu09b}, which includes
the comoving sound horizon at the drag epoch $r_s(z_d)$.
Note that $z_*$ is given by the fitting formula \cite{Hu96}:
\be
z_*=1048\, \left[1+ 0.00124 (\Omega_b h^2)^{-0.738}\right]\,
\left[1+g_1 (\Omega_m h^2)^{g_2} \right],
\ee
where
\ba
g_1 &= &\frac{0.0783\, (\Omega_b h^2)^{-0.238}}
{1+39.5\, (\Omega_b h^2)^{0.763}}\\
g_2 &= &\frac{0.560}{1+21.1\, (\Omega_b h^2)^{1.81}}
\ea
The redshift of the drag epoch $z_d$ is well approximated by 
\cite{EisenHu98}
\begin{equation}
z_d  =
 \frac{1291(\Omega_mh^2)^{0.251}}{1+0.659(\Omega_mh^2)^{0.828}}
\left[1+b_1(\Omega_bh^2)^{b2}\right],
\label{eq:zd}
\end{equation}
where
\begin{eqnarray}
  b_1 &= &0.313(\Omega_mh^2)^{-0.419}\left[1+0.607(\Omega_mh^2)^{0.674}\right],\\
  b_2 &= &0.238(\Omega_mh^2)^{0.223}.
\end{eqnarray}

CMB data are included in our analysis by adding
the following term to the $\chi^2$ of a given model
with $p_1=R(z_*)$, $p_2=l_a(z_*)$, $p_3=z_*$, and $p_4=r_s(z_d)$:
\be
\label{eq:chi2CMB}
\chi^2_{CMB}=\Delta p_i \left[ {\rm Cov}^{-1}_{CMB}(p_i,p_j)\right]
\Delta p_j,
\hskip .5cm
\Delta p_i= p_i - p_i^{data},
\ee
where $p_i^{data}$ are the maximum likelyhood values,
and ${\rm Cov}_{CMB}$ is the covariance matrix of 
[$R(z_*), l_a(z_*), z_*, r_s(z_d)]$ \cite{Komatsu09,Komatsu09b}.

\subsection{Baryon Acoustic Oscillation Data}
\label{sec:bao}

For the baryon acoustic oscillation (BAO) data, 
we use the measurement of $d_z \equiv r_s(z_d) /D_V(z)$ at
$z=0.2$ and $z=0.35$ by Percival et al. (2009) \cite{Percival09},
where
\be
D_V(z) \equiv \left[ \frac{ r(z)^2\, c z }{H(z)}\right]^{1/3}.
\ee
The inverse covariance matrix of ($d_{0.2}$, $d_{0.35}$), 
${\rm Cov}^{-1}_{BAO}$, is given by \cite{Percival09}: 
${\rm Cov}^{-1}_{BAO,11}=30124$,
${\rm Cov}^{-1}_{BAO,12}=-17227$, and 
${\rm Cov}^{-1}_{BAO,22}=86977$.

BAO data are included in our analysis by adding
the following term to the $\chi^2$ of a given model
with $p_1=d_{0.2}$ and $p_2=d_{0.35}$:
\be
\label{eq:chi2bao}
\chi^2_{BAO}=\Delta p_i \left[ {\rm Cov}^{-1}_{BAO}(p_i,p_j)\right]
\Delta p_j,
\hskip .5cm
\Delta p_i= p_i - p_i^{data},
\ee
where $p_1^{data}=d_{0.2}^{data}=0.1905$, and
$p_2^{data}=d_{0.35}^{data}=0.1097$ \cite{Percival09}.

\subsection{Gammay-ray Burst Data}
\label{sec:GRB}

We add gammay-ray burst (GRB) data to our analysis, since these are
complementary in redshift range to the SN Ia data.
We use GRB data in the form of the model-independent GRB distance 
measurements from Wang (2008b) \cite{Wang08b}, which were
derived from the data of 69 GRBs with $0.17 \le z \le 6.6$
from Schaefer (2007) \cite{Schaefer07}.

The GRB distance measurements are given in terms of \cite{Wang08b}
\be
\label{eq:rp}
\overline{r_p}(z_i)\equiv \frac{r_p(z)}{r_p(0.17)}, \hskip 1cm
r_p(z) \equiv \frac{(1+z)^{1/2}}{z}\, \frac{H_0}{ch}\, r(z),
\ee
where $r(z)$ is the comoving distance at $z$.

The GRB data are included in our analysis by adding
the following term to the $\chi^2$ of a given model:
\ba
\label{eq:rGRB1}
\chi^2_{GRB} &= & \left[\Delta \overline{r_p}(z_i)\right]  \cdot
\left(\mathrm{Cov}^{-1}_{GRB}\right)_{ij}\cdot
\left[\Delta \overline{r_p}(z_j)\right]
\nonumber\\
\Delta \overline{r_p}(z_i) &= & \overline{r_p}^{\mathrm{data}}(z_i)-\overline{r_p}(z_i),
\ea
where $\overline{r_p}(z)$ is defined by Eq.(\ref{eq:rp}).
The covariance matrix is given by
\be
\left(\mathrm{Cov}_{GRB}\right)_{ij}=
\sigma(\overline{r_p}(z_i)) \sigma(\overline{r_p}(z_j)) 
\left(\overline{\mathrm{Cov}}_{GRB}\right)_{ij},
\ee
where $\overline{\mathrm{Cov}}_{GRB}$ is the normalized covariance matrix
from Table 3 of Wang (2008b) \cite{Wang08b}, and
\ba
\label{eq:rGRB3}
\sigma(\overline{r_p}(z_i))  &= &\sigma\left(\overline{r_p}(z_i)\right)^+, \hskip 0.5cm \mathrm{if}\,\, 
\overline{r_p}(z) \ge \overline{r_p}(z)^{\mathrm{data}}; \nonumber\\
\sigma(\overline{r_p}(z_i))  &= &\sigma\left(\overline{r_p}(z_i)\right)^-, \hskip 0.5cm \mathrm{if}\,\, 
\overline{r_p}(z) < \overline{r_p}(z)^{\mathrm{data}},
\ea
where $\sigma\left(\overline{r_p}(z_i)\right)^+$ and 
$\sigma\left(\overline{r_p}(z_i)\right)^-$ are the 68\% C.L. errors
given in Table 2 of Wang (2008b) \cite{Wang08b}.

\subsection{Dark energy parametrization}
\label{sec:para}

Since we are ignorant of the the true nature of dark energy,
it is useful to measure the dark energy density function
$X(z)\equiv \rho_X(z)/\rho_X(0)$ as a free function of
redshift \cite{WangGarnavich,WangTegmark04,WangFreese06}.
Here we parametrize $X(z)$ by cubic-splining its values
at $z=0.25$, 0.5, 0.75, and 1.0, and assume that
$X(z>1)=X(z=1)$. Fixing $X(z>1)$ reflects the limit
of current data, and avoids making assumptions about early
dark energy that can be propagated into artificial constraints on
dark energy at low $z$ \cite{WangTegmark04,WangPia07}.

For comparison with the work of others, we also 
consider a dark energy equation of state linear in the cosmic scale 
factor $a$, $w_X(a)=w_0+(1-a)w_a$ \cite{Chev01}.
A related parametrization is \cite{Wang08a}
\ba
w_X(a) &= &\left(\frac{a_c-a}{a_c-1}\right) w_0
+\left(\frac{a-1}{a_c-1}\right) w_{0.5} \nonumber\\
 &= &\frac{a_cw_0-w_{0.5} + a(w_{0.5}-w_0)}{a_c-1}
\label{eq:wc}
\ea
with $a_c=2/3$ (i.e., $z_c=0.5$), and $w_{0.5}\equiv w_X(z=0.5)$.
Eq.(\ref{eq:wc}) is related to $w_X(z)=w_0+(1-a)w_a$  
by setting \cite{Wang08a}
\be
w_a= \frac{w_{0.5}-w_0}{1-a_c}, \hskip 1cm
{\rm or} \hskip 1cm
w_{0.5}=w_0+(1-a_c) w_a.
\label{eq:wa,wc}
\ee
Wang (2008a) \cite{Wang08a} showed that ($w_0$, $w_{0.5}$) are much less
correlated than ($w_0$, $w_a$), thus are a better set of parameters
to use.

\section{Results}

We perform a Markov Chain Monte Carlo (MCMC) likelihood analysis
\cite{Lewis02} to obtain ${\cal O}$($10^6$) samples for each set of 
results presented in this paper. 
We assum flat priors for all the parameters, and allow ranges 
of the parameters wide enough such that further increasing the allowed 
ranges has no impact on the results.
The chains typically have worst e-values (the
variance(mean)/mean(variance) of 1/2 chains)
much smaller than 0.005, indicating convergence.
The chains are subsequently 
appropriately thinned to ensure independent samples.

In addition to the SN Ia, CMB, BAO, and GRB data discussed in 
Sec.{\ref{sec:method}}, we impose a prior of 
$H_0 = 74.2 \pm 3.6\,$km$\,$s$^{-1}$Mpc$^{-1}$, from the Supernovae 
and $H_0$ for the Equation of State (SHOES) program on the HST \cite{Riess09}.

We do {\it not} assume a flat universe unless specifically noted.
In addition to the dark energy parameters described in Sec.\ref{sec:para},
we also constrain cosmological parameters ($\Omega_m, \Omega_k, h, \omega_b,
M_{SN}$), where $\omega_b\equiv \Omega_b h^2$, and $M_{SN}$ is used
to model the absolute distance scale of the SNe Ia.

\subsection{Model-Independent Distance Measurements from SNe Ia}

For an intuitive understanding of the SN Ia data, 
we parametrize a scaled distance
\be
r_p(z) \equiv \frac{r(z)}{cH_0^{-1} z} (1+z)^{0.41}
\label{eq:rp_sn}
\ee
by its values at $z_i=0.125i$, $i=1,2,...,10$, and $z_{11}=1.551$
(the highest redshift of SNe Ia in the ``Constitution''
and nearby+SDSS+ESSENCE+SNLS+HST data sets), and measure
$r_p(z_i)$ ($i=1,2,...,11$) from the SN Ia data.
The $r_p(z)$ at arbitrary $z$ is given by cubic spline interpolation,
thus {\it no} assumptions are made about cosmological models.
The $\{r_p(z_i)\}$ thus provide model-independent distance measurements
from SNe Ia.

Note that the scaled distances defined in Eqs.(\ref{eq:rp_sn}) and 
(\ref{eq:rp}) are similar, but with different choices of the 
power of $(1+z)$ used; this choice is made to make the scaled distance 
as flat as possible over the redshift range of interest, in order
to maximize the accuracy for cubic-spline. 

For cosmological models allowed by current data, the accuracy of interpolating 
$r_p(z)$ using $r_p(z_i)$ ($i=0,1,2,...,11$, with $z_0=0$ and $r_p(0)=1$) is 
around 0.4\%, compared to the exact distances.

Figs.{\ref{fig:rp}-{\ref{fig:rp_sdss} show the
distances measured from the ``Constitution''
and nearby+SDSS+ESSENCE+SNLS+HST data sets.
Clearly, the nearby+SDSS+ESSENCE+SNLS+HST set of 288 SNe Ia
gives {\it tighter} constraints on distances, although
it contains 109 fewer SNe Ia than the ``Constitution'' set.
\begin{figure} 
\psfig{file=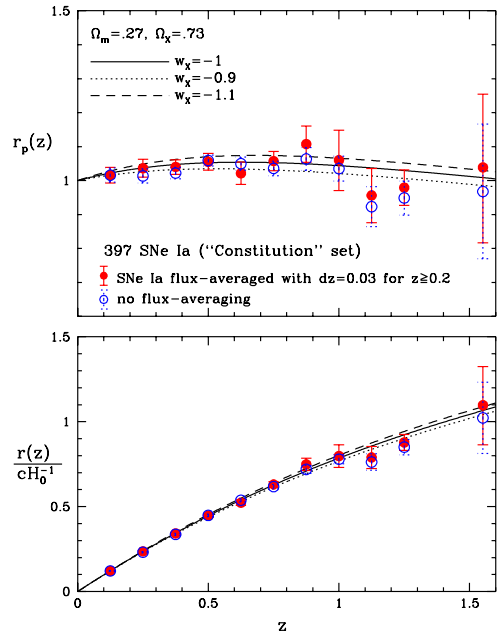,width=3.5in}\\
\caption[2]{\label{fig:rp}\footnotesize%
Distance measurements from the ``Constitution'' set of 397 SNe Ia.
and nearby+SDSS+ESSENCE+SNLS+HST data sets.
}
\end{figure}

\begin{figure} 
\psfig{file=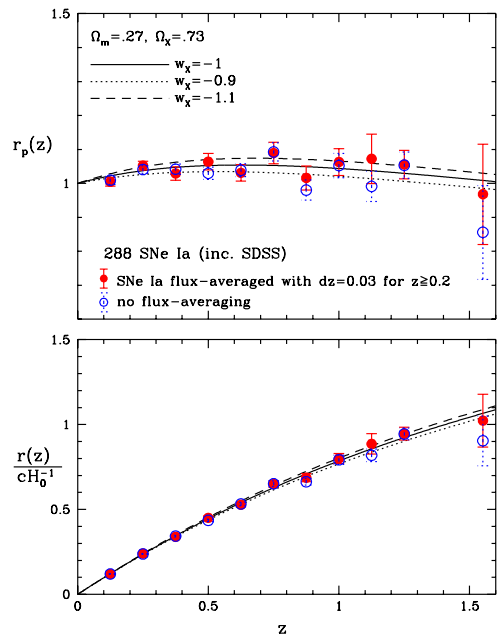,width=3.5in}\\
\caption[2]{\label{fig:rp_sdss}\footnotesize%
Distance measurements from the nearby+SDSS+ESSENCE+SNLS+HST
set of 288 SNe Ia.
}
\end{figure}

We have flux-averaged the SNe Ia at $z\ge 0.2$,
to minimize any likely systematic biases due to lensing or 
unknown systematic effects. It is interesting to note that
flux-averaging has a more significant effect on the
``Constitution'' set. This is as expected, since the
``Constitution'' set is less homogeneous than the
nearby+SDSS+ESSENCE+SNLS+HST data set, which include
103 SNe Ia from SDSS at $0.04 <z<0.42 $ \cite{Kessler09}.
Flux-averaging brings both data sets
closer to the prediction of a flat universe dominated
by a cosmological constant.

Table I lists the distances measured from the
nearby+SDSS+ESSENCE+SNLS+HST data set of 288 SNe Ia,
flux-averaged with $dz=0.03$ at $z\geq 0.2$.
Table II gives the corresponding normalized covariance matrix.
Note that the distances measured from SNe Ia,
$r_p(z_i)$ ($i=1,2,...,11$), are only moderately correlated.
Since the measurements of $r_p(z_i)$ ($i=1,2,...,11$), as given 
by Tables I and II, encode all the distance information from the 
nearby+SDSS+ESSENCE+SNLS+HST data set of 288 SNe Ia independent
of a cosmological model, they can be used in place of
the full data set following the prescription given
in Sec.\ref{sec:GRB}.

\begin{table*}[htb]
\caption{Distances measured from 288 SNe Ia with 68\% C.L. upper and lower uncertainties.}
\begin{center}
\begin{tabular}{|l|llccc|}
\hline
  & $z$  & $\overline{r_p}^{data}(z)$ & $\sigma(r_p)\,\,\,\,\,$  & $\sigma\left(r_p(z)\right)^+$  & 
 $\sigma\left(r_p(z)\right)^-$\\
 \hline
    0  &     0.0     &    1.0000   & --    &     --      &   -- \\
    1  & 0.125   & 1.0044  & 0.0129  &  0.9919  &  1.0171  \\
    2  & 0.25    & 1.0523  & 0.0129  &  1.0396  &  1.0651 \\    
    3  & 0.375   & 1.0292  & 0.0199  &  1.0095  &  1.0490\\    
    4  & 0.5     & 1.0638  & 0.0244  &  1.0397  &  1.0878 \\ 
    5  & 0.625   & 1.0311  & 0.0245  &  1.0067  &  1.0556\\  
    6  & 0.75    & 1.0895  & 0.0321  &  1.0575  &  1.1212\\  
    7  & 0.875   & 1.0160  & 0.0356  &  0.9807  &  1.0512\\  
    8  & 1.0     & 1.0629  & 0.0398  &  1.0234  &  1.1022 \\ 
    9  & 1.125   & 1.0725  & 0.0747  &  0.9996  &  1.1452\\  
   10  & 1.25    & 1.0557  & 0.0427  &  1.0136  &  1.0977 \\ 
   11  & 1.551   & 0.9682  & 0.1505  &  0.8203  &  1.1151 \\ 
\hline
\end{tabular}
\end{center}
\end{table*}

\begin{table*}[htb]
\caption{Normalized covariance matrix of distances measured from 288 SNe Ia}
\begin{center}
\begin{tabular}{|l|rrrrrrrrrrr|}
\hline
       & 1  & 2  & 3  &4  &5  &6 & 7 & 8& 9 & 10 & 11\\
      \hline
1&  1.0000 & 0.2539 & 0.5219 & 0.2016 & 0.3354 & 0.2148 & 0.2030 & 0.1741 & 0.0873 & 0.1728 & 0.0349\\
2&  0.2539 & 1.0000 & 0.1386 & 0.2120 & 0.1496 & 0.1483 & 0.1068 & 0.1192 & 0.0635 & 0.0971 & 0.0323\\
3&  0.5219 & 0.1386 & 1.0000 & 0.1288 & 0.2575 & 0.1413 & 0.1421 & 0.1130 & 0.0656 & 0.1106 & 0.0350\\
4&  0.2016 & 0.2120 & 0.1288 & 1.0000 &$-$0.0694 & 0.1473 & 0.0235 & 0.0917 & 0.0280 & 0.0650 & 0.0117\\
5&  0.3354 & 0.1496 & 0.2575 &$-$0.0694 & 1.0000 & 0.1236 & 0.1273 & 0.0677 & 0.0537 & 0.0809 & 0.0234\\
6&  0.2148 & 0.1483 & 0.1413 & 0.1473 & 0.1236 & 1.0000 &$-$0.1503 & 0.1709 &$-$0.0173 & 0.0521 &$-$0.0163\\
7&  0.2030 & 0.1068 & 0.1421 & 0.0235 & 0.1273 &$-$0.1503 & 1.0000 &$-$0.0877 & 0.0796 & 0.0609 & 0.0515\\
8&  0.1741 & 0.1192 & 0.1130 & 0.0917 & 0.0677 & 0.1709 &$-$0.0877 & 1.0000 & 0.0811 & 0.0299 &$-$0.0129\\
9&  0.0873 & 0.0635 & 0.0656 & 0.0280 & 0.0537 &$-$0.0173 & 0.0796 & 0.0811 & 1.0000 & 0.1779 & 0.3163\\
10& 0.1728 & 0.0971 & 0.1106 & 0.0650 & 0.0809 & 0.0521 & 0.0609 & 0.0299 & 0.1779 & 1.0000 &$-$0.4585\\
11& 0.0349 & 0.0323 & 0.0350 & 0.0117 & 0.0234 &$-$0.0163 & 0.0515 &$-$0.0129 & 0.3163 &$-$0.4585 & 1.0000\\
\hline
\end{tabular}
\end{center}
\end{table*}

\subsection{Measurements of the Dark Energy Density Function and $H(z)$}

The most transparent way to see how current data compare
to the prediction of the cosmological constant model is to 
measure the dark energy density function. 
Fig.{\ref{fig:rhoX}-{\ref{fig:rhoX_sdss} show the dark energy density function 
measured from combining SN Ia data with CMB, BAO, GRB data, and 
imposing the SHOES prior on $H_0$, for SNe Ia from the ``Constitution''
and the nearby+SDSS+ESSENCE+SNLS+HST data sets respectively, without 
assuming a flat universe. Fig.{\ref{fig:rhoXflat} and 
Fig.{\ref{fig:rhoXflat_sdss} are similar to Fig.{\ref{fig:rhoX} 
and Fig.{\ref{fig:rhoX_sdss}, except a flat universe is assumed.

\begin{figure} 
\psfig{file=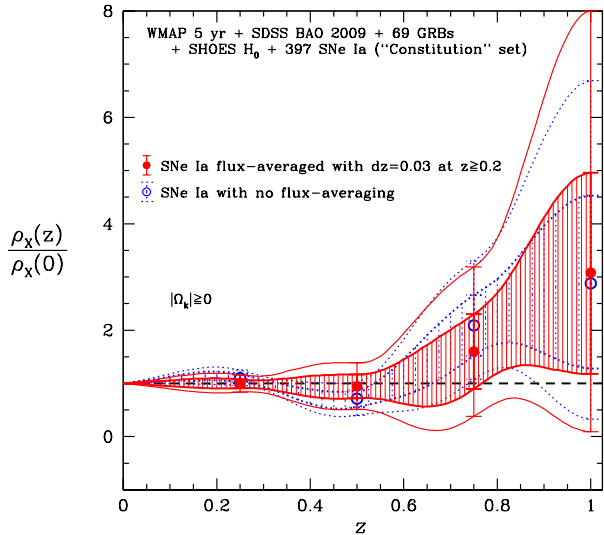,width=3.5in}\\
\vspace{-0.4in}
\caption{\label{fig:rhoX}\footnotesize%
Dark energy density function $X(z)\equiv \rho_X(z)/\rho_X(0)$
measured from combining SN Ia data (the ``Constitution'' set
of 397 SNe Ia) with CMB, BAO, GRB data, 
and imposing the SHOES prior on $H_0$.
The 68\% (shaded) and 95\% confidence level regions are shown.
}
\end{figure}

\begin{figure} 
\psfig{file=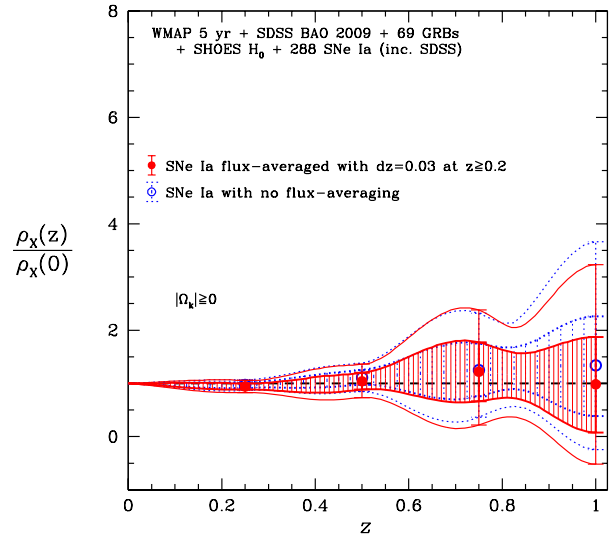,width=3.5in}\\
\vspace{-0.4in}
\caption{\label{fig:rhoX_sdss}\footnotesize%
Dark energy density function $X(z)\equiv \rho_X(z)/\rho_X(0)$
measured from combining SN Ia data (nearby+SDSS+ESSENCE+SNLS+HST data 
set of 288 SNe Ia) with CMB, BAO, and GRB data,
and imposing the SHOES prior on $H_0$.
The 68\% (shaded) and 95\% confidence level regions are shown.
}
\end{figure}

\begin{figure} 
\psfig{file=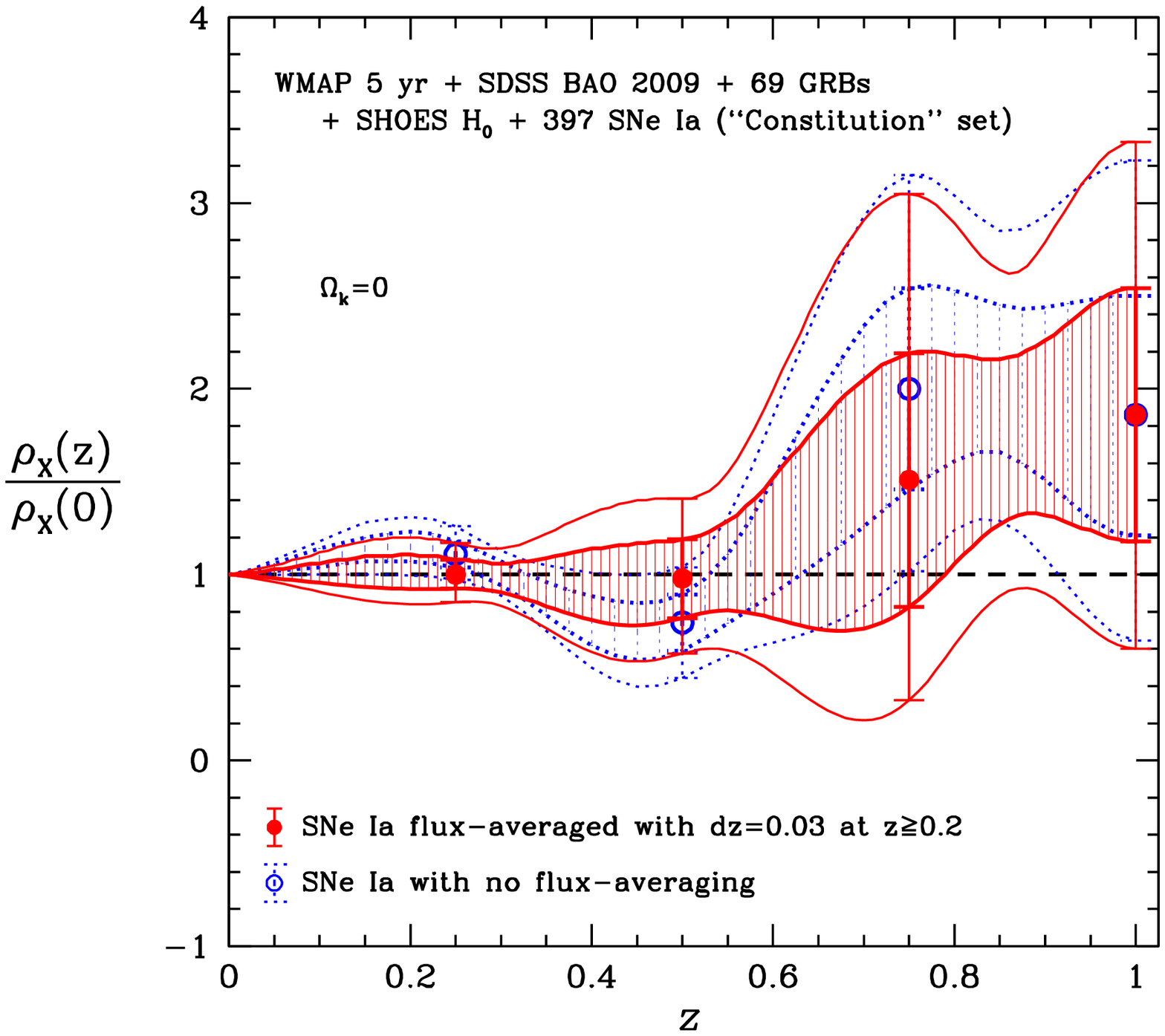,width=3.5in}\\
\vspace{-0.4in}
\caption{\label{fig:rhoXflat}\footnotesize%
Dark energy density function $X(z)\equiv \rho_X(z)/\rho_X(0)$
measured from combining SN Ia data (the ``Constitution'' set
of 397 SNe Ia) with CMB, BAO, GRB data, 
and imposing the SHOES prior on $H_0$.
The 68\% (shaded) and 95\% confidence level regions are shown.
A flat universe is assumed.
}
\end{figure}

\begin{figure} 
\psfig{file=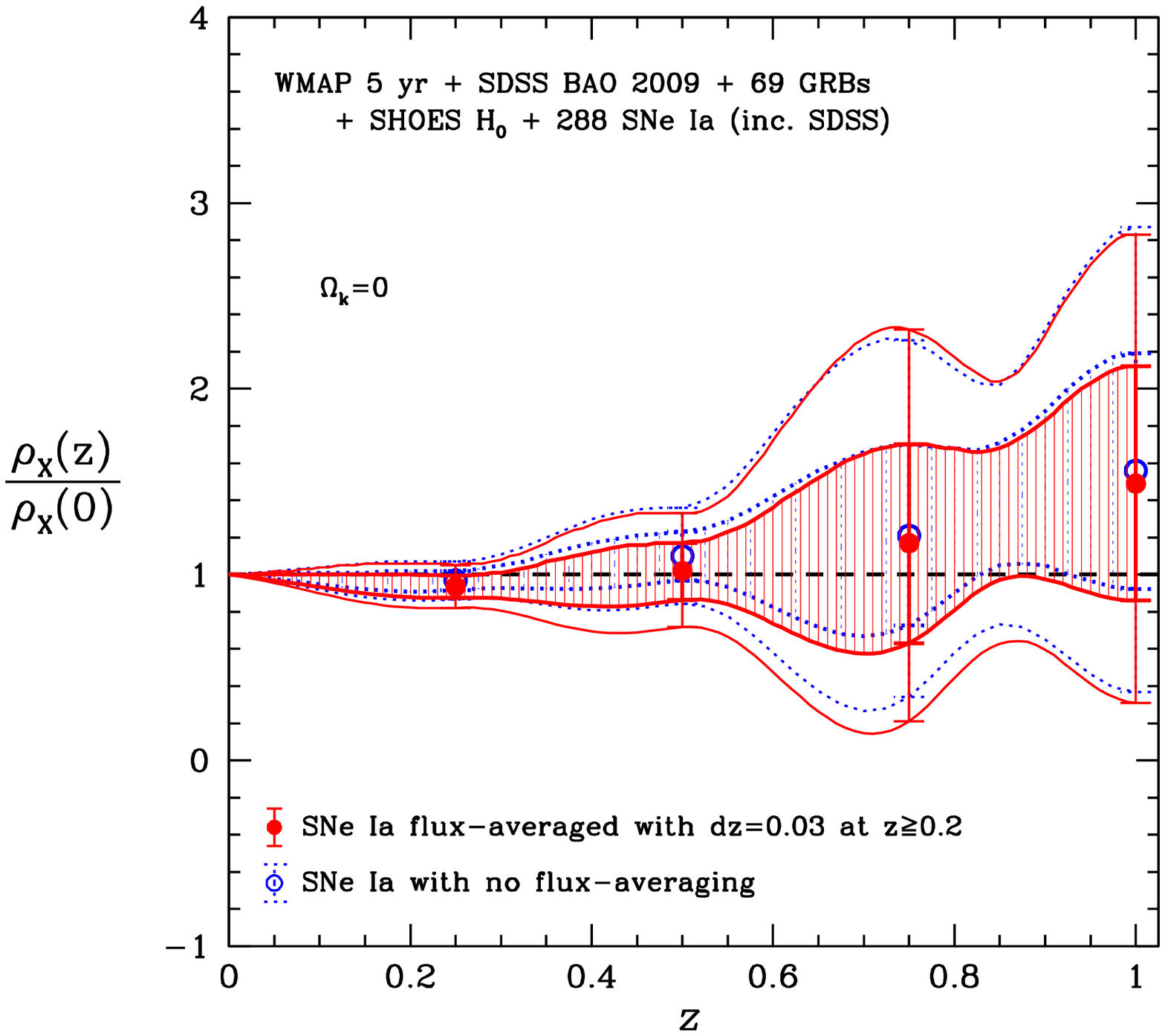,width=3.5in}\\
\psfig{file=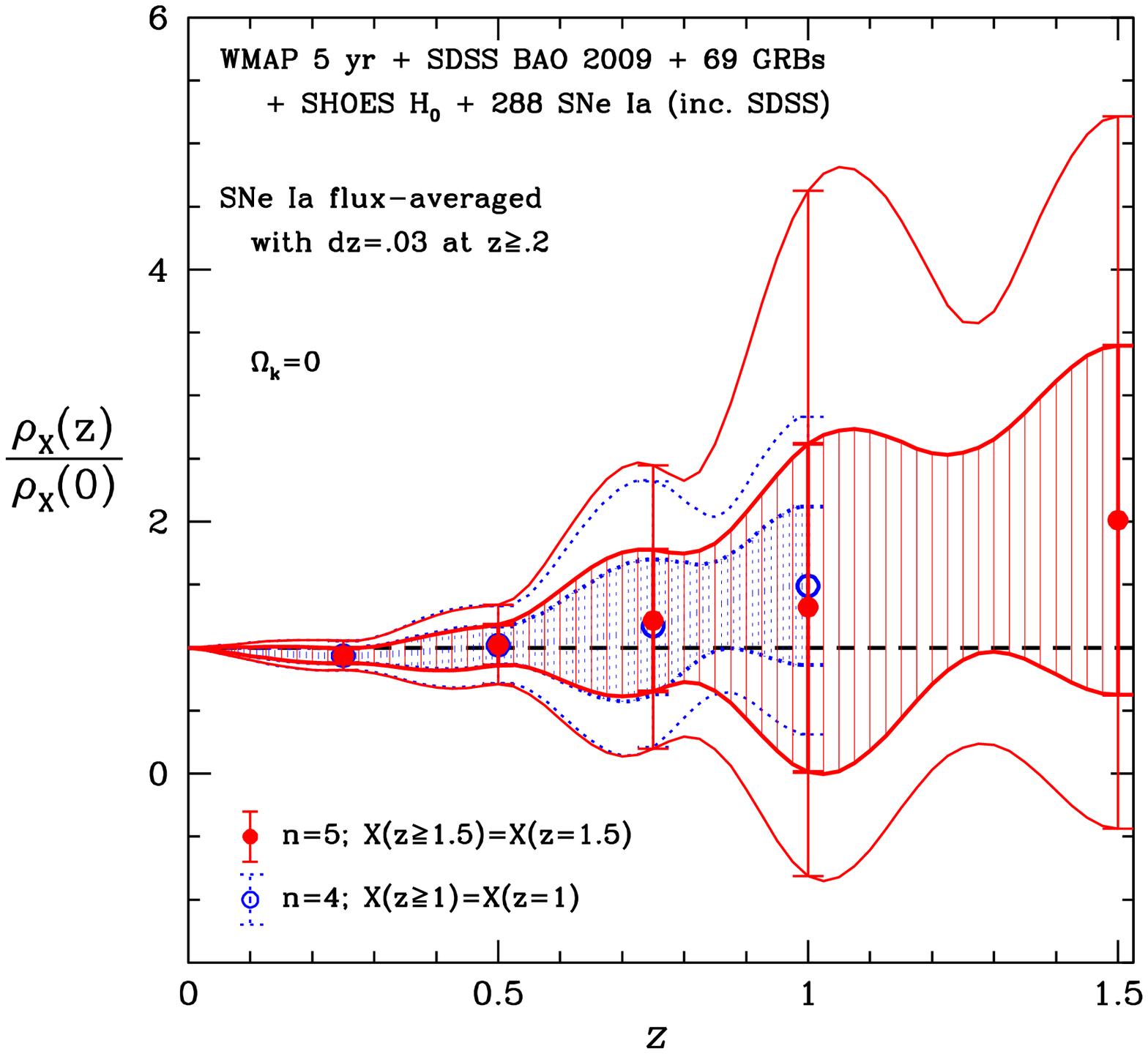,width=3.5in}\\
\vspace{-0.4in}
\caption{\label{fig:rhoXflat_sdss}\footnotesize%
Dark energy density function $X(z)\equiv \rho_X(z)/\rho_X(0)$
measured from combining SN Ia data (nearby+SDSS+ESSENCE+SNLS+HST data 
set of 288 SNe Ia) with CMB, BAO, and GRB data,
and imposing the SHOES prior on $H_0$.
The 68\% (shaded) and 95\% confidence level regions are shown.
A flat universe is assumed.
(a) Assuming $X(z\geq 1)=X(z=1)$.
(b) Same as (a), but with $X(z=1.5)$ added as a parameter,
and assuming $X(z\geq 1.5)=X(z=1.5)$.
}
\end{figure}

Again, the same trend as in Figs.{\ref{fig:rp}}-{\ref{fig:rp_sdss}}
is seen: using the nearby+SDSS+ESSENCE+SNLS+HST data set of SNe Ia gives 
much more stringent constraints on dark energy than using the ``Constitution''
set of SNe Ia, and gives measurements that are closer to a
cosmological constant. Flux-averaging again has larger impact
on the results from using the ``Constitution'' set of SNe Ia, and brings 
the measurements closer to that predicted by a cosmological constant.

Note that the $X(z)$ measured using the ``Constitution'' set of SNe Ia
actually deviates from a cosmological constant at $\sim$2$\sigma$
{\it without} flux-averaging; the downturn in the measured $X(z)$
at $z\sim 0.25-0.5$ is consistent with that found by 
Huang et al. (2009) \cite{Huang09}.
Since this apparent variation in $X(z)$ disappears in
the results from using the nearby+SDSS+ESSENCE+SNLS+HST set of SNe Ia
{\it without} flux-averaging, it is likely that it originated
from unknown systematic effects in the ``Constitution'' set of SNe Ia.

Fig.\ref{fig:rhoXflat_sdss}(a) shows that assuming a flat universe,
dark energy is detected at $>2\sigma$ at $z=1$ using the 
nearby+SDSS+ESSENCE+SNLS+HST data set of 288 SNe Ia, together 
with CMB, BAO, and GRB data. Fig.\ref{fig:rhoXflat_sdss}(b) is 
the same as Fig.\ref{fig:rhoXflat_sdss}(a), but with 
$X(z=1.5)$ added as a parameter, and assuming $X(z\geq 1.5)=X(z=1.5)$.
The $X(z)$ results at $z\leq 0.75$ remain about the same, while the 
errors for $X(z=1)$ are larger (but the mean value of $X(z)$ remains
about the same) because of the additional parameter allowed. 
Independent of the assumptions about $X(z)$ at $z\geq 1$,
dark energy is detected at $> 98$\% confidence level for $z\leq 0.75$.

\begin{figure} 
\psfig{file=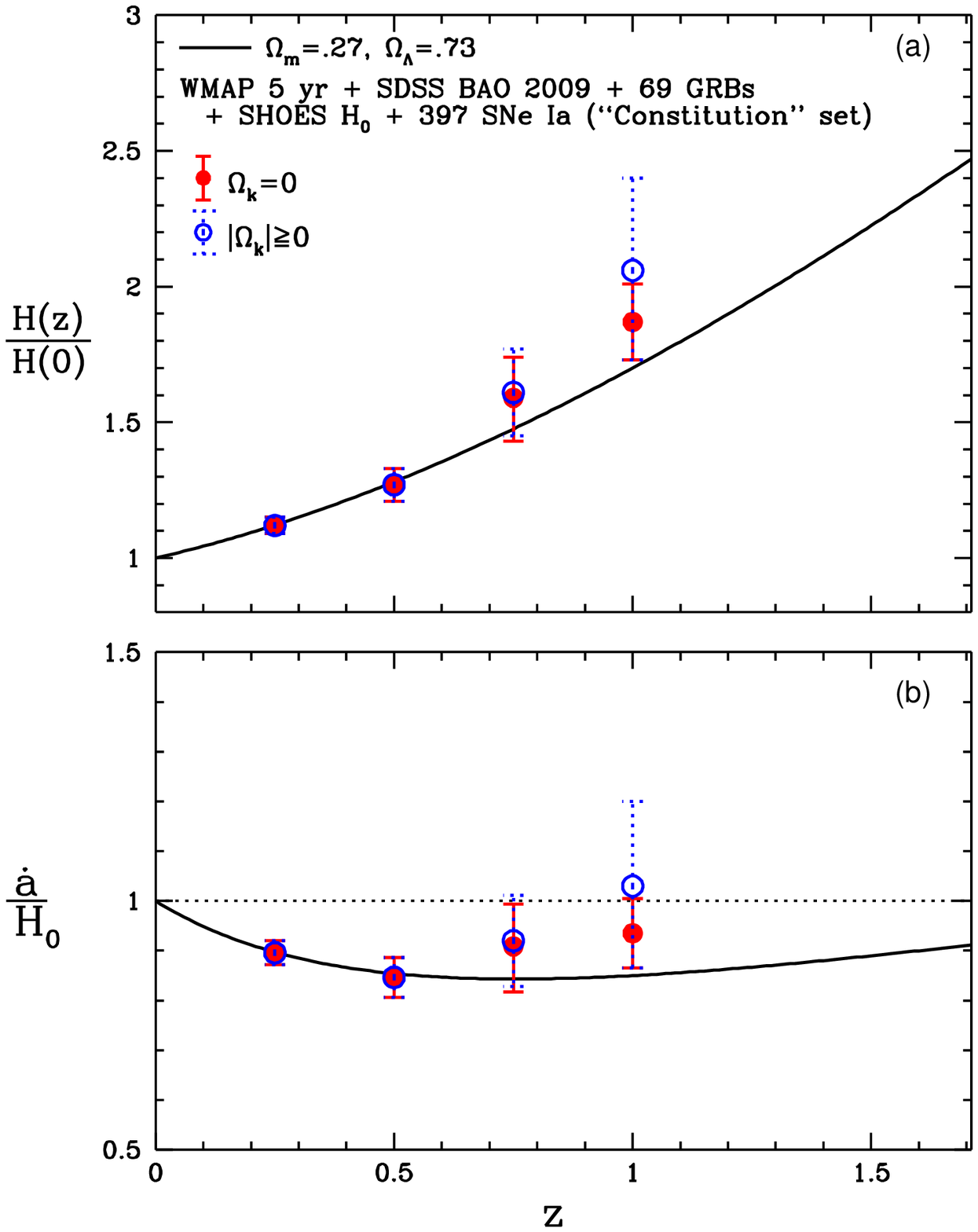,width=3.5in}\\
\caption{\label{fig:Hz}\footnotesize%
The measured cosmic expansion history $H(z)$ corresponding to
Fig.{\ref{fig:rhoX}}. The error bars represent the 68\% confidence level.
}
\end{figure}

\begin{figure} 
\psfig{file=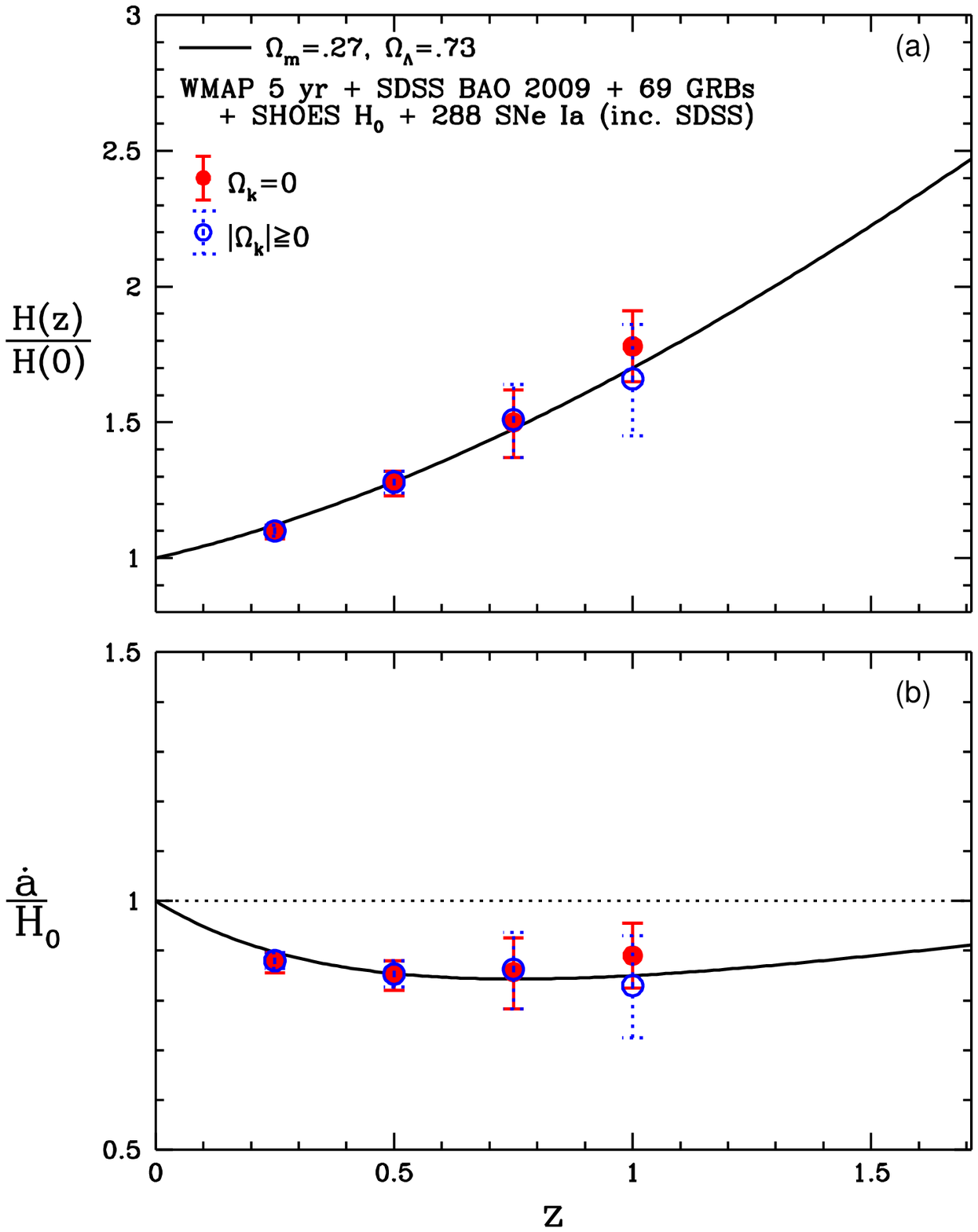,width=3.5in}\\
\caption{\label{fig:Hz_sdss}\footnotesize%
The measured cosmic expansion history $H(z)$ corresponding to
Fig.{\ref{fig:rhoX_sdss}}. The error bars represent the 68\% confidence level.
}
\end{figure}

Figs.{\ref{fig:Hz}-{\ref{fig:Hz_sdss} are the measurements of
the cosmic expansion history $H(z)$ that correspond to
Fig.{\ref{fig:rhoX}-{\ref{fig:rhoX_sdss}.

Table III gives the dark energy density function $X(z)\equiv \rho_X(z)/\rho_X(0)$
and the cosmic expansion history $H(z)$ measured from current data
(nearby+SDSS+ESSENCE+SNLS+HST data set of 288 SNe Ia, together
with CMB, BAO, GRB data, and imposing the SHOES prior on $H_0$). 
The $H(z)$ measurements are derived using
Eq.(\ref{eq:H(z)}). Tables IV and V give the normalized covariance
matrices of the $X(z)$ and $H(z)$ measurements. Note that both the 
$X(z)$ and $H(z)$ measurements are only weakly correlated.

\begin{table*}[htb]
\caption{$X(z)$, $H(z)$, and cosmological parameters
estimated from current data with 68\% C.L. upper and lower uncertainties.}
\begin{center}
\begin{tabular}{|l|llcc|}
\hline 
 & $\mu$ & $\sigma$ & $\sigma^{-}$ & $\sigma^+$ \\
 \hline 
$X(z=0.2)$ & 0.942 & 0.059 & 0.884 & 1.001  \\   
$ X(z=0.5)$ & 1.041 & 0.159 & 0.883 & 1.199 \\   
$X(z=0.75)$ & 1.223 & 0.556 & 0.667 & 1.776 \\  
$ X(z=1.0)$ & 0.987 & 0.960 & 0.089 & 1.869 \\  
$\Omega_m$ & 0.262 & 0.016 & 0.246 & 0.278 \\    
$\Omega_k$ &$-$0.009 & 0.012 &$-$0.020 & 0.003 \\   
$h$        & 0.709 & 0.020 & 0.690 & 0.729 \\    
$\omega_b$ & 0.02377 & 0.00062 & 0.02321 & 0.02441 \\   
\hline
$H(z=0.25)$ & 1.10  & 0.021 & 1.08 & 1.12  \\   
$H(z=0.5)$ &  1.28  & 0.043 & 1.24 & 1.32 \\    
$H(z=0.75)$ & 1.51  & 0.132 & 1.37 & 1.64 \\    
$H(z=1.0)$ &  1.66  & 0.215 & 1.45 & 1.86 \\    
\hline
\end{tabular}
\end{center}
\end{table*}

\begin{table*}[htb]
\caption{Normalized covariance matrix of $X(z)$ from current data}
\begin{center}
\begin{tabular}{|l|rrrr|}
\hline
       & 1  & 2  & 3  &4  \\
    \hline
1&  1.0000   & $-$0.1273 & $-$0.0668 & 0.0078\\
2& $-$0.1273 & 1.0000    &$-$0.0374  &$-$0.2789\\
3&  $-$0.0668 & $-$0.0374 &  1.0000   &$-$0.3638\\
4&   0.0078  & $-$0.2789  & $-$0.3638 & 1.0000\\
\hline
\end{tabular}
\end{center}
\end{table*}

\begin{table*}[htb]
\caption{Normalized covariance matrix of $H(z)$ from current data}
\begin{center}
\begin{tabular}{|l|rrrr|}
\hline
       & 1  & 2  & 3  &4  \\
    \hline
1&    1.0000 &$-$0.1665 &$-$0.1212 & 0.1493\\
2& $-$0.1665  &1.0000 &$-$0.1524 &$-$0.2281\\
3& $-$0.1212 &$-$0.1524 & 1.0000 &$-$0.3605\\
4&  0.1493 &$-$0.2281& $-$0.3605 & 1.0000\\
\hline
\end{tabular}
\end{center}
\end{table*}

\subsection{Constraints on ($w_0,w_a)$, ($w_0,w_{0.5}$)}

For comparison with the work of others, 
Figs.{\ref{fig:09_w0wc}-{\ref{fig:09_w0wc_sdss}
show the 68\% and 95\% joint confidence level contours of 
($w_0,w_a)$ and ($w_0,w_{0.5}$),
from combining SN Ia data with CMB, BAO, GRB data, and 
imposing the SHOES prior on $H_0$, for SNe Ia from the ``Constitution''
and the nearby+SDSS+ESSENCE+SNLS+HST data sets respectively,

\begin{figure} 
\psfig{file=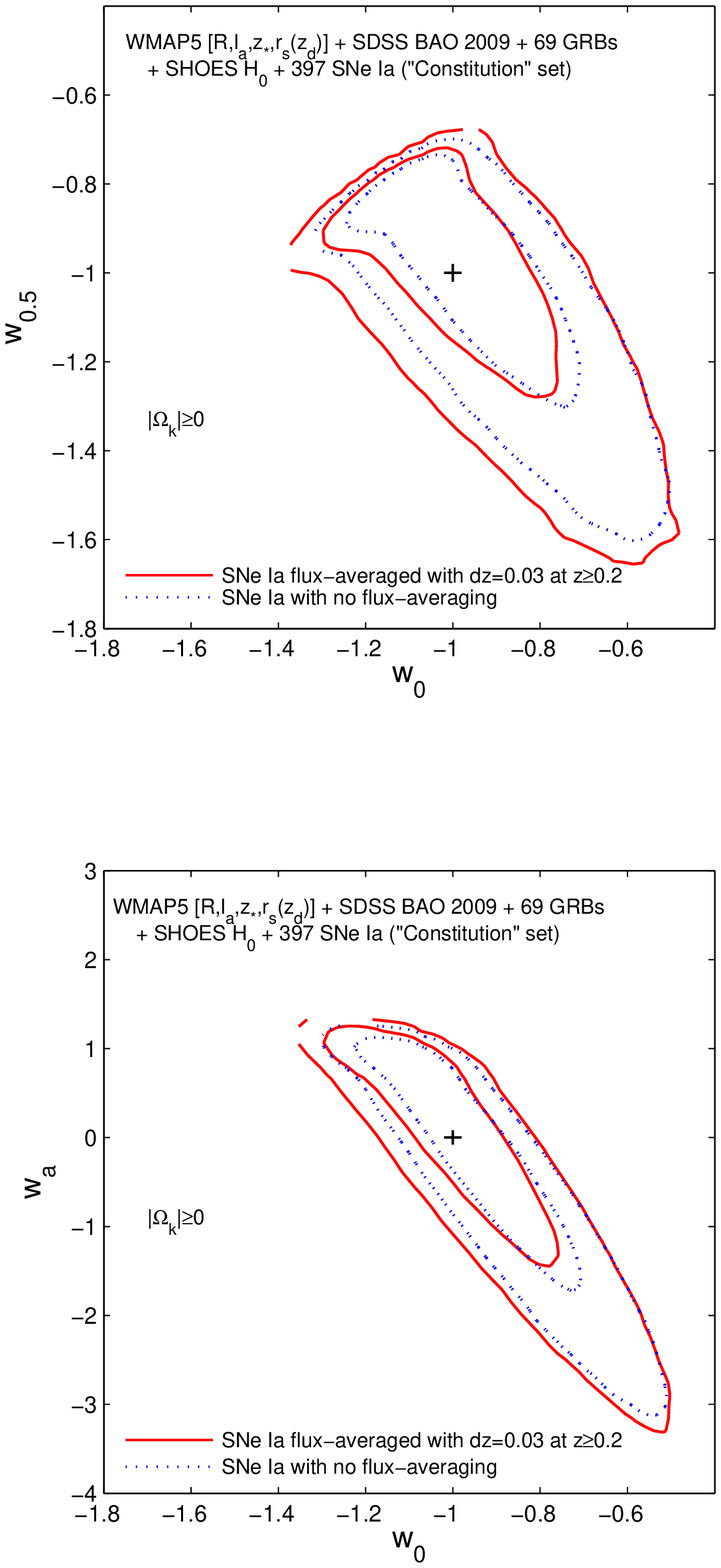,width=2.8in}\\
\caption{\label{fig:09_w0wc}\footnotesize%
The 68\% and 95\% joint confidence level contours on constraints on ($w_0,w_a)$ and 
($w0,w_{0.5}$), using SN Ia data (the ``Constitution'' set
of 397 SNe Ia) together with CMB, BAO, GRB data,
and imposing the SHOES prior on $H_0$.
}
\end{figure}

\begin{figure} 
\psfig{file=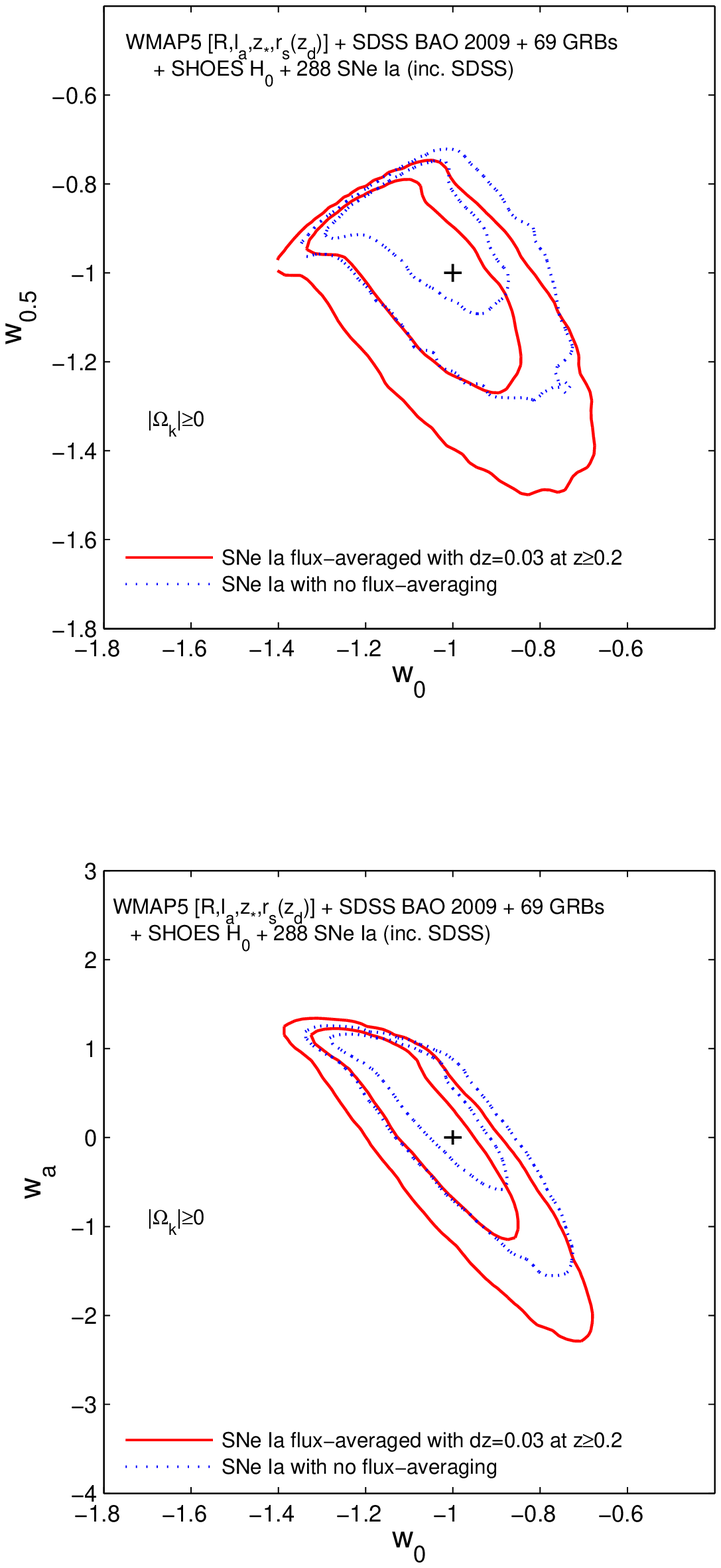,width=2.8in}\\
\caption{\label{fig:09_w0wc_sdss}\footnotesize%
The 68\% and 95\% joint confidence level contours on constraints on ($w_0,w_a)$ and 
($w0,w_{0.5}$), using SN Ia data (the nearby+SDSS+ESSENCE+SNLS+HST 
set of 288 SNe Ia) together with CMB, BAO, GRB data,
and imposing the SHOES prior on $H_0$.
}
\end{figure}

It is interesting to note that flux-averaging has a {\it larger}
impact on the ($w_0,w_a)$ and ($w_0,w_{0.5}$) error contours
from the combined data using the nearby+SDSS+ESSENCE+SNLS+HST
set of SNe Ia, compared to the combined data using the
``Constitution'' of SNe Ia. This is in contrast to what
we found when measuring the dark energy density function $X(z)$
at $z=0.25, 0.5, 0.75, 1$ (see Figs.\ref{fig:rhoX}-\ref{fig:rhoX_sdss}). 
A comparison of the correlation coefficients in Tables IV and VI reveal 
that assuming a linear dark energy equation-of-state results in much 
stronger correlations among the dark energy parameters. This obsures 
some of the information about dark energy contained in the data.

\subsection{Dark Energy Figure of Merit from Current Data}

Wang (2008a) \cite{Wang08a} defined a general dark energy Figure-of-Merit 
(FoM)
\be
{\rm FoM} = \frac{1}{\sqrt{{\rm det Cov}(f_1, f_2, ...f_N)}}
\label{eq:FoM}
\ee
where $(f_1, f_2, ...f_N)$ is the set of parameters that have been
chosen to parametrize dark energy. 

The Dark Energy Task Force (DETF) defined the dark energy FoM to be 
the inverse of the area enclosed by the 95\% confidence level contour 
of $(w_0,w_a)$ \cite{detf}. The areas enclosed by contours are difficult 
to calculate for real data, as these contours can be quite irregular (see 
Figs.{\ref{fig:09_w0wc}-{\ref{fig:09_w0wc_sdss}).
The definition of Eq.(\ref{eq:FoM}) has the advantage of being
easy to calculate for either real or simulated data.
For $(f_1, f_2)=(w_0,w_a)$, ${\rm FoM}$ of Eq.(\ref{eq:FoM})
is proportional to the FoM defined by the DETF for ideal Gaussian-distributed
data, and the same as the relative FoM used by the DETF in
Fisher matrix forecasts.

Table VI shows the dark energy FoM from SN Ia data 
together with CMB, BAO, GRB data, and imposing the SHOES prior 
on $H_0$, for the dark energy density function 
$X(z)\equiv \rho_X(z)/ \rho_X(0)$ measured
at $z=0.25, 0.5, 0.75, 1.0$ (with X(z) given by cubic spline
elsewhere and assuming $X(z>1)=X(1)$, see Sec.\ref{sec:para})
and a dark energy equation-of-state linear in $a$ parametrized
by ($w_0,w_{0.5}$) and ($w_0,w_a$) respectively.

\begin{table*}[htb]
\caption{Dark energy FoM from current data}
\begin{center}
\begin{tabular}{|l|c|cccc|cccc|}
\hline
data: CMB+BAO+GRB+ & FoM$_r(\{X_i\})$ & $\sigma(w_0)$ & $\sigma(w_{0.5})$ & 
$r_{w_0,w_{0.5}}$ & FoM$_r(w_0,w_{0.5})$ &
$\sigma(w_0)$ & $\sigma(w_a)$ & $r_{w_0,w_a}$ &FoM$_r(w_0,w_a)$\\
\hline
288 SNe Ia (inc. SDSS) & 230.7 & 0.1473 & 0.1688& $-$0.6676 &54.0&
0.1475  &0.8708& $-$0.9029& 18.1\\
397 SNe Ia (``Constitution'') & 44.1 & 0.1841  & 0.2252 & $-$0.7530 &36.6 &
0.1845 &  1.1349& $-$0.9232& 12.4\\
\hline
\end{tabular}
\end{center}
\end{table*}

It is interesting to note that for $\{X(z_i)\}$ ($z_i$=0.25, 0.5, 0.75, 1.0),
the factor of improvement in the dark energy FoM is 5.2 when the 
nearby+SDSS+ESSENCE+SNLS+HST data set of 288 SNe Ia
is used instead of the ``Constitution'' set of 397 SNe Ia
(see Table VI); this quantifies the difference between 
Fig.{\ref{fig:rhoX_sdss}} and Fig.{\ref{fig:rhoX}}.
For a linear dark energy equation of state, the improvement
in FoM is a little less than $\sim$50\% (see Table VI).
The same equation of state linear in $a$ can be written in either 
($w_0,w_{0.5}$) or ($w_0,w_a$) (see Sec.\ref{sec:para}).
As expected, the parameters ($w_0,w_{0.5}$) are significantly less correlated
than ($w_0,w_a$) \cite{Wang08a}, thus represent a better choice
of parameters. The improvement in the dark energy FoM is
slightly larger for ($w_0,w_{0.5}$) than for ($w_0,w_a$).

\subsection{The Impact of Adding CMB Constraints on $r_s(z_d)$}

Compared to previous work using CMB distances priors,
we have added the constraints on $r_s(z_d)$. 
Not surprisingly, this has negligible impact on
$X(z)$ measured from current data, since we have assumed
that $X(z>1)=X(z=1)$.

Figs.{\ref{fig:09_w0wc_rsd}}-{\ref{fig:09_w0wa_rsd_sdss}} show
the difference in the marginalized probability density distributions
of the dark energy and cosmological parameters
using [$R(z_*), l_a(z_*), z_*, r_s(z_d)]$ (as we have done
throughout this paper), and using [$R(z_*), l_a(z_*), \omega_b$].
Clearly, adding CMB constraints on $r_s(z_d)$ tightens the
constraints on ($w_0,w_a)$ and ($w_0,w_{0.5}$), as these
are parameters from assuming a dark energy equation-of-state
linear in the cosmic scale factor $a$, thus imply strong
assumptions about early dark energy that are propagated 
to the low and intermediate redshifts \cite{WangTegmark04}.

We find that the constraints on [$R(z_*), l_a(z_*), z_*, r_s(z_d)]$ 
summarize all useful information from CMB data that are
relevant to dark energy and are geometric and independent 
of dark energy perturbations.
Further adding constraints on $z_d$ does {\it not} add new
information, since $z_d$ is well approximated by Eq.(\ref{eq:zd}),
which only depends on $\Omega_m h^2$ and $\Omega_b h^2$,
and which in turn are already constrained by [$R(z_*), l_a(z_*), z_*, r_s(z_d)]$.

\begin{figure} 
\psfig{file=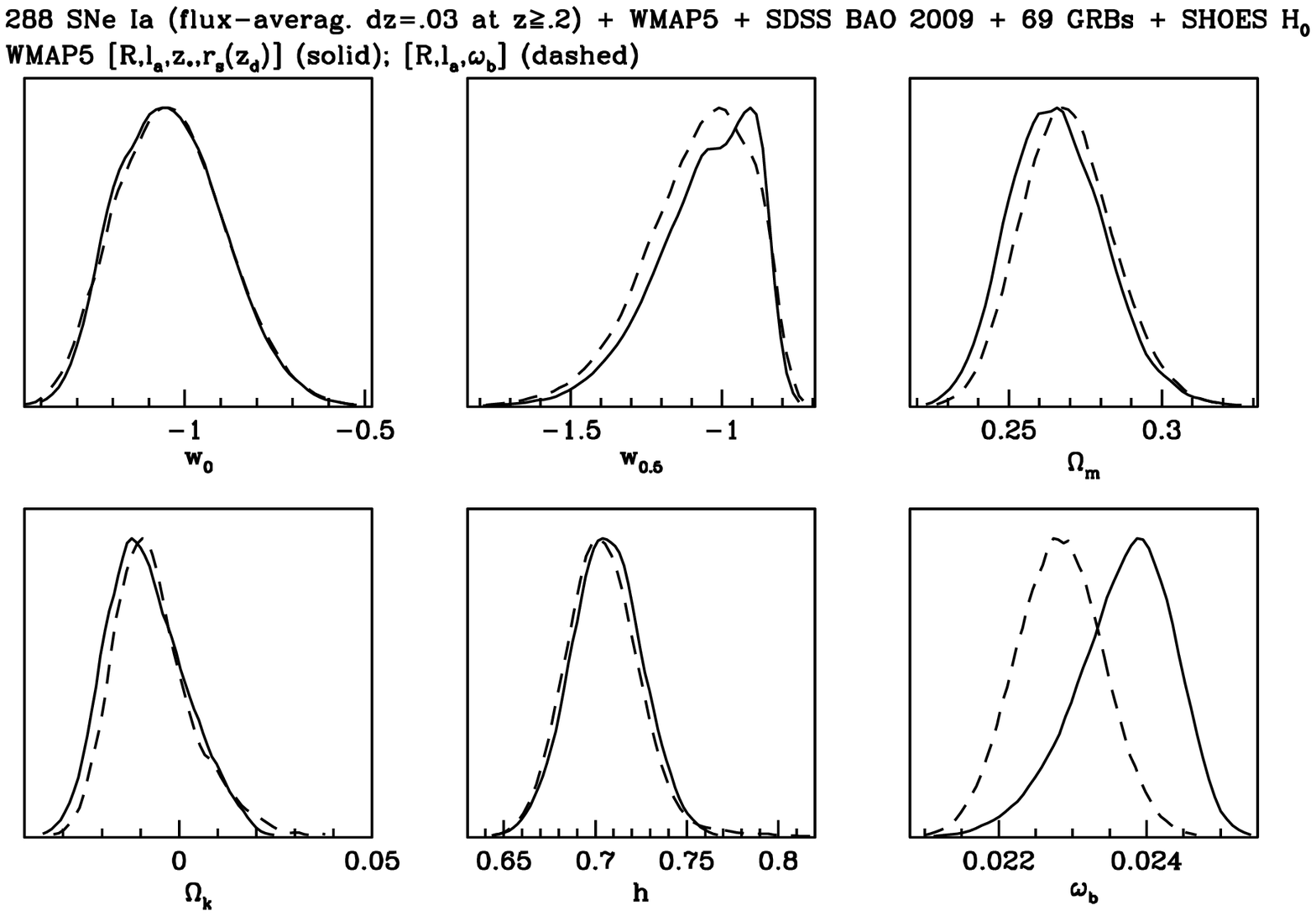,width=3.5in}\\
\vspace{-1in}
\caption[2]{\label{fig:09_w0wc_rsd}\footnotesize%
Marginalized probability distributions of the dark energy and cosmological 
parameters using SN Ia data (the nearby+SDSS+ESSENCE+SNLS+HST 
set of 288 SNe Ia flux-averaged at $z\geq 0.2$) together with CMB, 
BAO, GRB data, and imposing the SHOES prior on $H_0$,
assuming a linear dark energy equation
of state parametrized by ($w_0,w_{0.5}$) (see Eq.[\ref{eq:wc}]).
The solid and dashed lines indicate the use of 
the CMB data summarized by [$R(z_*), l_a(z_*), z_*, r_s(z_d)]$,
and [$R(z_*), l_a(z_*), \omega_b$] respectively.
}
\end{figure}

\begin{figure} 
\psfig{file=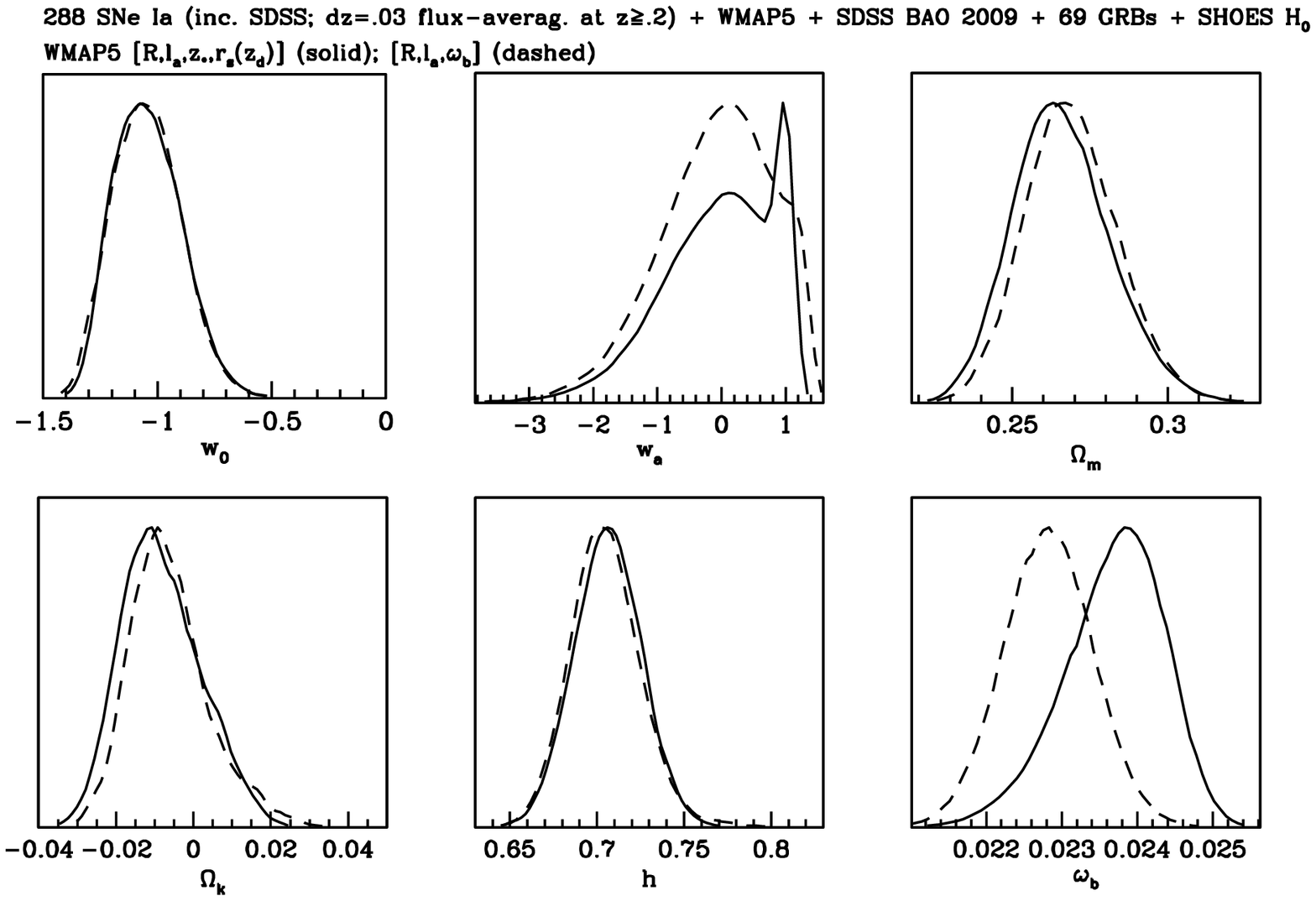,width=3.5in}\\
\vspace{-1in}
\caption[2]{\label{fig:09_w0wa_rsd_sdss}\footnotesize%
Same as Fig.{\ref{fig:09_w0wc_rsd}}, except the linear dark energy
equation of state is parametrized by the usual ($w_0,w_a$).
}
\end{figure}

\section{Summary and Discussion}

It is likely that current sets of SNe Ia may be contaminated
by unknown systematic effects \cite{Qi09,Sanchez09}.
We have used the flux-averaging of SNe Ia (which was developed
to reduce the weak lensing systematic effect \cite{Wang00b,WangPia04,Wang05}) 
to help reduce the impact of unknown systematic effects.
Since the SNe Ia in each redshift bin are flux-averaged and 
{\it not} used directly in the likelihood analysis, the systematic
bias that results from unknown systematic errors from individual
SNe Ia can be minimized. For homogeneous data without unknown
systematic errors, flux-averaging should not change the results
qualitatively at intermediate redshifts (where the lensing
effect is negligible). Thus in addition to reducing lensing and 
lensing-like systematic effects, flux-averaging of SN Ia data provide 
a useful test for the presence of unknown systematic effects
at intermediate redshifts.\footnote{Note that flux-averaging corrects 
the bias in estimated distances due to weak lensing;
this makes the statistical errors bigger when the number of SNe Ia 
flux-averaged in a bin is small.}

We find that flux-averaging has a significantly smaller effect on
the nearby+SDSS+ESSENCE+SNLS+HST data set of 288 SNe Ia \cite{Kessler09}, 
compared to the ``Constitution'' set of 397 SNe Ia \cite{Hicken09},
see Figs.\ref{fig:rp} and \ref{fig:rp_sdss}.
This same trend continues when the SN Ia data are combined
with CMB, BAO, GRB data, and the SHOES measurement of $H_0$
to measure the dark energy density function, see  
Figs.\ref{fig:rhoX}-\ref{fig:rhoXflat_sdss}.
Furthermore, flux-averaging brings both data sets closer to
the predictions of a flat universe dominated by a cosmological constant
(see Figs.\ref{fig:rp}-\ref{fig:Hz_sdss}).
The nearby+SDSS+ESSENCE+SNLS+HST data set of SNe Ia (together
with CMB, BAO, GRB data, and the SHOES measurement of $H_0$)
are consistent with a cosmological constant at 68\% confidence
level with or without flux-averaging of SNe Ia (see Fig.\ref{fig:rhoX_sdss}).
Without flux-averaging, the combined data using the ``Constitution'' set 
of 397 SNe Ia seem to indicate a deviation from a cosmological constant at
$\sim$95\% confidence at $0\la z \la 0.8$, but is consistent with a 
cosmological constant at $\sim$68\% confidence when SNe Ia are flux-averaged 
(see Fig.\ref{fig:rhoX}).

It is interesting to note that for a dark energy density function $X(z)$
parametrized by $\{X(z_i)\}$ ($z_i$=0.25, 0.5, 0.75, 1.0)
and assumed to be constant at $z>1$, the factor of improvement in the dark 
energy FoM is 5.2 when the nearby+SDSS+ESSENCE+SNLS+HST data set of 288 SNe Ia
is used instead of the ``Constitution'' set of 397 SNe Ia (see Table VI).
For a linear dark energy equation of state, the improvement
in FoM is a little less than $\sim$50\%. 
This indicates that current data already contain more information
about dark energy than can be adequately represented by
a linear dark energy equation of state. 
Using the nearby+SDSS+ESSENCE+SNLS+HST data set of 288 SNe Ia
gives more stringent constraints on dark energy than using
the ``Constitution'' set of 397 SNe Ia, because the former data set
includes 103 SNe Ia from SDSS at $0.04 <z<0.42 $ \cite{Kessler09}
which are not in the latter data set, and because the latter 
data set is less homogeneous and likely more prone to unknown 
systematic effects.

We have measured the dark energy density function $X(z)$
from data without imposing $X(z) \geq 0$. 
Since the probability density distributions for $X(z)$ 
that we have obtained from the current data extend to 
$X(z)<0$, it is not possible to convert the measurements
of $X(z)$ into measurements of $w_X(z)=p_X(z)/\rho_X(z)$,
or the deceleration parameter $q(z)$ (which depends on $w_X(z)$).
$X(z)$ is related to the dark energy equation of state $w_X(z)$ 
as follows \cite{WangGarnavich}:
\begin{equation}
\label{eq:rhoprimew}
X(z)\equiv\frac{ \rho_X(z)}{\rho_X(0)} = 
\exp\left\{ \int_0^z {\rm d}z'\, \frac{3[1+w_X(z')]}{1+z'} \right\}. 
\end{equation} 
{\it Hence parametrizing dark energy with $w_X(z)$ 
implicitly assumes that $\rho_X(z)$ does not change sign 
in cosmic time (i.e., $X(z)\geq 0$).} 
This precludes whole classes of dark energy models in which $\rho_X(z)$ 
becomes negative in the future (``Big Crunch'' models, see 
Refs.\cite{Linde87,WangLinde04} for an example) 
\cite{WangTegmark04}. Thus the measurement of $X(z)$, without
assuming $X(z) \geq 0$, contains more information about dark 
energy than that of $w_X(z)$ or $q(z)$.

Our work complements Genovese et al. (2009) \cite{Genovese09},
and Bogdanos \& Nesseris (2009) \cite{Bogdanos09}, both of these
papers assumed a flat universe, and analyzed SN Ia data only.
Genovese et al. (2009) used model-testing (see e.g. \cite{Liddle06})
to diffentiate assumptions about $w_X(z)$, and 
derived parametric and non-parametric estimators of $w_X(z)$ \cite{Genovese09}.
Bogdanos \& Nesseris (2009) used genetic algorithms to derive
model-independent constraints on the distance modulus from SNe Ia,
which are then used to constrain $w_X(z)$ \cite{Bogdanos09}.
In this paper, we have derived model-independent constraints
on $X(z)$, instead of $w_X(z)$, for both a flat universe and
a universe with arbitrary spatial curvature.
We also use newer SN Ia data (which significantly tighten the
dark energy constraints), and combine with CMB, BAO, GRB data, 
and the SHOES measurement of $H_0$.

Finally, we note that the ($w_0,w_a$) FoM for current data
(the nearby+SDSS+ESSENCE+SNLS+HST data set of 288 SNe Ia flux-averaged,
with CMB, BAO, GRB data, and the SHOES prior on $H_0$) is 18.1,
more than a factor of two larger than that found by Wang (2008a)
using data available in March 2008 (182 SNe Ia compiled by 
Riess et al. 2007 \cite{Riess07}, $[R(z_*), l_a(z_*), \Omega_b h^2]$ 
from the five year WMAP observations, the SDSS measurement of the 
BAO scale by Eisenstein et al. 2005 \cite{Eisen05}, 
and assuming the HST prior of $H_0=72\pm 8\,$km$\,$s$^{-1}$Mpc$^{-1}$).
Most of this gain results from the significantly improved distance 
measurements from SNe Ia (see Figs.\ref{fig:rp}-\ref{fig:rp_sdss}).
Assuming a flat universe, dark energy is detected at
$> 98$\% confidence level for $z\leq 0.75$ using the combined data
with 288 SNe Ia from nearby+SDSS+ESSENCE+SNLS+HST,
independent of the assumptions about $X(z\geq 1)$
(see Fig.\ref{fig:rhoXflat_sdss}).

In order to make solid progress on probing dark energy, it will be
essential to launch aggressive observational projects to obtain
large and uniform sets of SNe Ia spanning the redshift range of
$0\la z \la 2 $ \cite{Wang00a}.

\bigskip

{\bf Acknowledgements}
I am grateful to Eiichiro Komatsu for helpful comments and
for making the covariance matrix for [$R(z_*), l_a(z_*), \Omega_b h^2, r_s(z_d)$] 
from WMAP 5 year data available, and I acknowledge the use of cosmomc
in processing the MCMC chains.


\begin{thebibliography}{}

\bibitem[Riess et al.~(1998)]{Riess98}
Riess, A. G, {\etal}, 1998, Astron. J., 116, 1009

\bibitem[Perlmutter et al.~(1999)]{Perl99} 
Perlmutter, S. {\etal}, 1999, ApJ, 517, 565

\bibitem[Komatsu et al.(2009)]{Komatsu09}
Komatsu, E., et al. 2009, Astrophys.J.Suppl., 180, 330

\bibitem{Sollerman09}
Sollerman, J., et al., Astrophysical Journal 703 (2009) 1374

\bibitem{Serra09}
Serra, P., et al., arXiv:0908.3186

\bibitem{Reid09}
Reid, B.A., et al., arXiv:0907.1659

\bibitem{Biswas09}
Biswas, R., Wandelt, B.D., arXiv:0903.2532

\bibitem{Daly09}
Daly, R.A., et al., ApJ, 691, 1058 (2009)

\bibitem{Samushia09}
Samushia, L., et al., arXiv:0906.2734

\bibitem{Shafieloo09}
Shafieloo, A., Sahni, V., Starobinsky, A.A., arXiv:0903.5141 

\bibitem[Freese et al.(1987)]{Freese87}
Freese, K., Adams, F.C., Frieman, J.A.,
 and Mottola, E.,
Nucl. Phys. {\bf B287}, 797 (1987).

\bibitem[Linde(1987)]{Linde87}
Linde A D, ``Inflation And Quantum Cosmology,'' in
{\it Three hundred years of gravitation}, (Eds.: Hawking, S.W. and Israel, W.,
Cambridge Univ. Press, 1987), 604-630.

\bibitem[Peebles \& Ratra(1988)]{Peebles88}
Peebles, P.J.E., and Ratra, B., 1988, ApJ, 325, L17

\bibitem[Wetterich(1988)]{Wett88}
Wetterich, C., 1988, Nucl.Phys., B302, 668 

\bibitem[Frieman et al.(1995)]{Frieman95}
Frieman, J.A., Hill, C.T., Stebbins, A., and Waga, I., 1995, PRL, 75, 2077 

\bibitem[Caldwell, Dave \& Steinhardt(1998)]{Caldwell98}
Caldwell, R., Dave, R., \& Steinhardt, P.J., 1998, PRL, 80, 1582

\bibitem{Kaloper06}
Kaloper, N., \& Sorbo, L., JCAP 0604 (2006) 007

\bibitem{Chiba09}
Chiba, T.; Dutta, S., \& Scherrer, R.J., Phys.Rev.D80:043517,2009

\bibitem[Sahni \& Habib(1998)]{SH98}
Sahni, V., \& Habib, S., 1998, PRL, 81, 1766 

\bibitem[Parker \& Raval(1999)]{Parker99}  
Parker, L., and Raval, A., 1999, PRD, 60, 063512

\bibitem[Boisseau et al.(2000)]{Boisseau00}
Boisseau, B., Esposito-Far\`ese, G., 
Polarski, D. \& Starobinsky, A. A. 2000, Phys. Rev. Lett., 85, 2236

\bibitem[Dvali, Gabadadze, \& Porrati(2000)]{DGP00}
Dvali, G., Gabadadze, G., \& Porrati, M. 2000,
Phys.Lett. B485, 208

\bibitem[Freese \& Lewis(2002)]{Freese02}
Freese, K., \& Lewis, M., 2002, Phys. Lett. B, 540, 1 

\bibitem{Pad08}
Padmanabhan, T., arXiv:0807.2356

\bibitem{Kahya09}
Kahya, E. O.; Onemli, V. K.; Woodard, R. P., arXiv:0904.4811

\bibitem{OCallaghan09}
O'Callaghan, E., Gregory, R., Pourtsidou, A., JCAP 0909:020,2009

\bibitem{Copeland06}
Copeland, E.~J., Sami, M., Tsujikawa, S., IJMPD, 15 (2006), 1753

\bibitem{Ruiz07}
Ruiz-Lapuente, P., Class. Quantum. Grav., 24 (2007), 91 

\bibitem{Ratra07}
Ratra, B., Vogeley, M.~S., arXiv:0706.1565 (2007)

\bibitem{Frieman08}
Frieman, J., Turner, M., Huterer, D., ARAA, 46, 385 (2008)

\bibitem{Caldwell09}
Caldwell, R. R., \& Kamionkowski, M., arXiv:0903.0866

\bibitem{Uzan09}
Uzan, J.-P., arXiv:0908.2243

\bibitem{Cimatti09}
Cimatti, A., et al., Experimental Astronomy, 23, 39 (2009)

\bibitem{Refregier09}
Refregier, A., et a., Exper.Astron.23:17-37,2009

\bibitem{Hicken09}
Hicken, M., et al. 2009, ApJ, 700, 1097

\bibitem{Kessler09}
Kessler, R., et al. 2009, arXiv0908.4274

\bibitem[Guy et al.(2005)]{Guy05}
Guy, J.; Astier, P.; Nobili, S.; Regnault, N.; Pain, R.,
2005, A\&A,443, 781	

\bibitem[Guy et al.(2007)]{Guy07}
Guy, J., et al.	2007, A\&A, 466, 11	

\bibitem[Phillips(1993)]{Phillips93}
Phillips, M. M. 1993, ApJ, 413, L105 

\bibitem[Riess, Press, \& Kirshner(1995)]{Riess95}
Riess, A.G., Press, W.H., \& Kirshner, R.P., ApJ, 438, L17 (1995)

\bibitem[Jha, Riess, \& Kirshner(2007)]{Jha07}
Jha, S.; Riess, A.G.; Kirshner, R.P. 2007, ApJ, 659, 122	

\bibitem{Qi09}
Qi, S.; Lu, T.; \& Wang, F.-Y., MNRAS, 398 (2009) L78-L82

\bibitem{Sanchez09}
Bueno Sanchez, J.C.; Nesseris, S., \& Perivolaropoulos,  L., arXiv:0908.2636

\bibitem[Wang(2000b)]{Wang00b}
Wang, Y., ApJ 536, 531 (2000b)

\bibitem[Wang \& Mukherjee(2004)]{WangPia04}
Wang, Y., \& Mukherjee, P. 2004, ApJ, 606, 654

\bibitem[Wang(2005)]{Wang05}
Wang, Y., JCAP, 03, 005 (2005), astro-ph/0406635

\bibitem[Wang \& Mukherjee(2007)]{WangPia07}
Wang, Y., \& Mukherjee, P., PRD, 76, 103533 (2007)

\bibitem[Page(2003)]{Page03}
Page, L., et al. 2003, ApJS, 148, 233 

\bibitem{Komatsu09b}
Komatsu, E., ``wmap$_{-}$prior$_{-}$for$_{-}$bao.pdf'' at
http://gyudon.as.utexas.edu/$\sim$komatsu/wmap5/

\bibitem{Wang08a}
Wang, Y., 2008a, Phys. Rev. D 77, 123525 

	
\bibitem[Hu \& Sugiyama(1996)]{Hu96}
Hu, W., \& Sugiyama, N. 1996, ApJ, 471, 542

\bibitem[Eisenstein \& Hu(1998)]{EisenHu98}
Eisenstein, D. \& Hu, W. 1998, ApJ, 496, 605

\bibitem{Percival09}
Percival, W.J., et al. 2009, arXiv0907.1660

\bibitem{Wang08b}
Wang, Y., 2008b, PRD, 78, 123532 

\bibitem{Schaefer07}
Schaefer, B. E., 2007, ApJ, 660, 16

\bibitem[Wang \& Garnavich(2001)]{WangGarnavich}
Wang, Y., and Garnavich, P. 2001, ApJ, 552, 445

\bibitem[Wang \& Tegmark(2004)]{WangTegmark04}
Wang, Y., \& Tegmark, M. 2004, Phys. Rev. Lett., 92, 241302 

\bibitem[Wang \& Freese(2006)]{WangFreese06}
Wang, Y., \& Freese, K. 2006, Phys.Lett. B632, 449
(astro-ph/0402208)

\bibitem[Chev01(2001)]{Chev01}
Chevallier, M., \& Polarski, D. 2001, Int. J. Mod. Phys. D10,
213

\bibitem[Lewis02(2002)]{Lewis02}
Lewis, A., \& Bridle, S. 2002, PRD, 66, 103511

\bibitem{Riess09}
Riess, A.G. , et al. 2009, ApJ, 699, 539

\bibitem{Huang09}
Huang, Q.-G., Li, M.; Li, X.-D.; Wang, S., arXiv:0905.0797 

\bibitem[deft(2006)]{detf}
Albrecht, A.; Bernstein, G.; Cahn, R.; Freedman, W. L.; Hewitt, J.;
Hu, W.; Huth, J.; Kamionkowski, M.; Kolb, E.W.; Knox, L.; Mather, J.C.;
Staggs, S.; Suntzeff, N.B., Report of the Dark Energy Task Force, 
astro-ph/0609591 

\bibitem{WangLinde04}
Wang, Y.; Kratochvil, J. M.; Linde, A.; \& Shmakova, M., 
JCAP 0412 (2004) 006

\bibitem{Genovese09}
Genovese, C., et al., Annals of Applied Statistics 2009, 
Vol. 3, No. 1, 144

\bibitem{Bogdanos09}
Bogdanos, C., \& Nesseris, S., JCAP05(2009)006

\bibitem{Liddle06}
Liddle, A.R.; Mukherjee, P.; Parkinson, D.; \& Wang, Y.,
Phys.Rev.D74 (2006) 123506 

\bibitem[Riess(2007)]{Riess07}
Riess, A.G., et al., astro-ph/0611572

\bibitem[Eisenstein et al.(2005)]{Eisen05}
Eisenstein, D., et al., ApJ, 633, 560


\bibitem[Wang(2000a)]{Wang00a}
Wang, Y., ApJ 531, 676 (2000a)


\end{thebibliography}
\end{document}